\documentclass[twocolumn]{webofc}
\newcommand{\bea}{\begin{eqnarray}}
\newcommand{\eea}{\end{eqnarray}}
\begin{document}
\title{Happy Birthday, Ultra-Cold Neutron!\thanks{This article covers the music part of a lecture given by H.A. at the International Workshop on Particle Physics at Neutron Sources, PPNS, in Grenoble/France in May 2018 with a composition on the occasion of the 50th anniversary of the first production of ultra-cold neutrons.}}
%
% subtitle is optionnal
%
%%%\subtitle{Do you have a subtitle?\\ If so, write it here}

\author{\firstname{Hartmut} \lastname{Abele}\inst{1}\fnsep\thanks{\email{abele@ati.ac.at}}\and
\firstname{Tobias} \lastname{Jenke}\inst{2} \and
\firstname{Hartmut} \lastname{Lemmel}\inst{1,2}
        % etc.
}

\institute{Technische Universit\"at Wien, Atominstitut, Stadionallee 2, 1020 WIEN, Austria
\and Institut Laue-Langevin, 71 avenue des Martyrs,
CS 20156, 38042 GRENOBLE Cedex 9, France}

\abstract{%
  What is driving the accelerated expansion of the universe and do we have an alternative for Einstein's cosmological constant? What is dark matter made of? Do extra dimensions of space and time exist? Is there a preferred frame in the universe? To which extent  is left-handedness a preferred symmetry in nature? What's the origin of the baryon asymmetry in the universe? These fundamental and open questions are addressed by precision experiments using ultra-cold neutrons. This year, we celebrate the 50th anniversary of their first production, followed by first pioneering experiments. Actually, ultra-cold neutrons were discovered twice in the same year – once in the eastern and once in the western world~\cite{Lushchikov:1969,Steyerl:1969}. For five decades now research projects with ultra-cold neutrons have contributed to the determination of the force constants of nature’s fundamental interactions, and several technological breakthroughs in precision allow to address the open questions by putting them to experimental test. To mark the event and tribute to this fabulous object, we present a birthday song for ultra-cold neutrons with acoustic resonant transitions~\cite{Cronenberg:2018}, which are based solely on properties of ultra-cold neutrons, the inertial and gravitational mass of the neutron~$m$, Planck's constant~$h$, and the local gravity~$g$. We make use of a musical intonation system that bears no relation to basic notation and basic musical theory as applied and used elsewhere~\cite{Riemann:1978} but addresses two fundamental problems of music theory, the problem of reference for the concert pitch and the problem of intonation.

%We celebrate the 50$^{th}$ anniversary of an experimental breakthrough, the first production of ultra-cold neutrons, facilitating experiments with unprecedented sensitivity using quantum interference techniques. To mark the event, we present a birthday song for these neutrons. The song "Happy Birthday" is recomposed in such a way that the musical notation system refers solely to properties of the neutron, the inertial and gravitational mass $m_n$, Planck’s constant $\hbar$, and the acceleration of the Earth $g$. Such a quantum mechanically founded music bears little relation to standard music notation as applied and used elsewhere, but addresses two fundamental problems
%of music theory, which we will discuss in some detail.
}
\maketitle
\section{A tribute to a fabulous object}
\label{intro}
What is an ultra-cold neutron? Following a pragmatic definition~\cite{Steyerl:1977}, such a neutron is able to become reflected from a given surface under any angle of incidence.
%capable of undergoing total external reflection in
%a given material at any angle of incidence. %even at normal incidence  or magnetic device
%become reflected from a given surface under any angle of incidence.
%an ultra-cold neutron is a neutron that reflects under all angles with respect to a given surface.
This feature allows us to store neutrons in a box or in a so-called neutron bottle. As a consequence, an observation time of up to few times its lifetime is manageable, thus enabling highly sensitive experiments. The ultra-cold neutron became thus a tool and an object for precision experiments. The search for a permanent electric dipole moment of the neutron investigates a high-energy scale in particle physics that cannot be reached by accelerators on Earth. The present experimental limit on this quantity is $|d_n|$ <3.0$\times$10$^{-26}$\,$\textnormal{e\,cm} \,\,\mathrm{(90
\% \, C.L.)}$~\cite{Pendlebury:2015}. At the current level of
sensitivity,  energy changes down to 10$^{-22}\,$eV can
be detected. A common technique
uses the Ramsey resonance method of separate oscillating fields~\cite{Ramsey:1949},
where polarized neutrons precess in a magnetic
field. The
precession frequency will change in the presence of an electric
field if the neutron has an electric dipole moment. The measurements are
made with ultra-cold neutrons stored in a cell permeated by uniform electric and
magnetic
fields~\cite{Altarev:1978se,Altarev:1981,Pendlebury:1984,Altarev:1986,Smith:1990,Altarev:1992,Altarev:1996,Harris:1999,Baker:2006,Pendlebury:2015}. This search investigates CP violating mechanisms beyond the Standard Model.
 These are necessary to explain the matter-antimatter asymmetry in our universe. A byproduct is a search for axionlike dark matter through nuclear spin precession.  "This null result sets the first laboratory constraints on the coupling of axion dark matter to gluons, which improve on astrophysical limits by up to 3 orders of magnitude"~\cite{Abel:2017}.

This is just an example how neutrons contribute to key issues that have emerged in particle physics and cosmology and are expected to be decisive for unanswered fundamental questions. Addressed by physics with neutrons as a tool~\cite{Dubbers:2011,ABELE:2008} are basic questions like "What is driving the accelerated expansion of the universe and do we have an alternative for Einstein's cosmological constant?", "What is dark matter made of?", "Do extra dimensions of space and time exist?", "Is there a preferred frame in the universe?", "To what extent is left-handedness a preferred symmetry in nature?", "What's the origin of the baryon asymmetry in the universe?", and so on.

\begin{figure*}
\centering
\includegraphics[width=14.cm,clip]{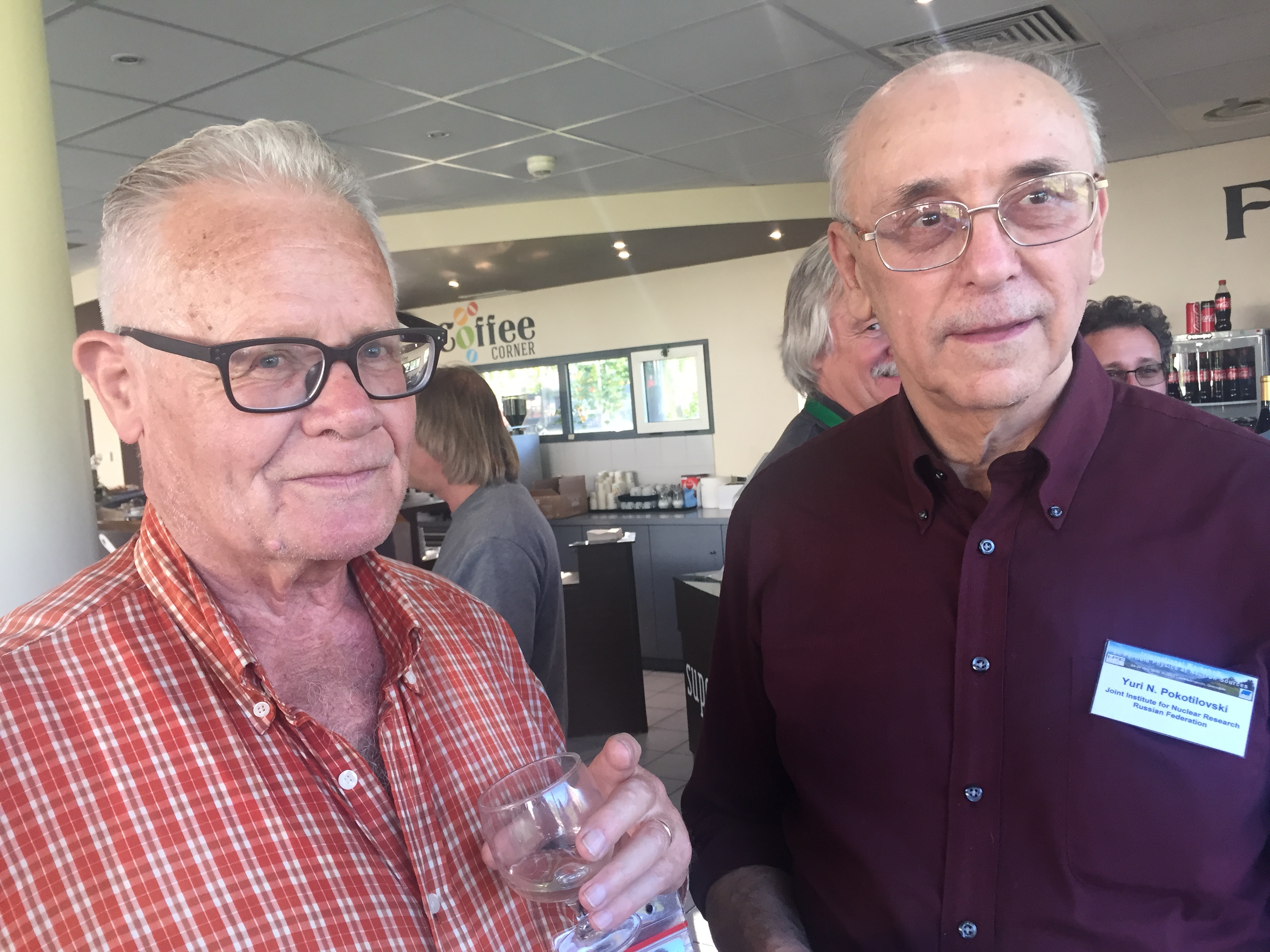}
 \caption{This photograph taken at the PPNS workshop in May 2018 shows Yuri N. Pokotilovskii, member of the Dubna team (right) and Albert Steyerl (left), two of the pioneers of research with ultra-cold neutrons. }
\label{fig-2a}       % Give a unique label
\end{figure*}
With neutrons serving as an object basic properties have been studied. Examples include the neutron lifetime~\cite{Mampe:1989a,Arzumanov:2000,Serebrov:2008,Ezhov:2009,Pichlmaier:2010,Serebrov:2018,Pattie:2018,Ezhov:2018} and other decay parameters like $\beta$-decay correlation coefficients~\cite{Pattie:2009,Mendenhall:2013,Brown:2018}, measurements of its magnetic moment, quantum mechanical~\cite{Luschikov:1978,Rauch:2015} or neutron optical~\cite{Frank:2006,Frank:2011} properties, and searches for a charge of the neutron~\cite{Durstberger-Rennhofer:2011}. The $\beta$-decay measurements are complemented by experiments with cold neutrons for the lifetime~\cite{Byrne:1996,Dewey:2003,Nico:2005,Yue:2013}, see also the review~\cite{Wietfeldt:2011}, and measurements of correlation coefficients~\cite{Bopp:1986,Schreckenbach:1995,Abele:1997,Yerozolimsky:1997,Serebrov:1998,Abele:2002,Soldner:2004,Kreuz:2005,Schumann:2007,Schumann:2008,Chupp:2012,Kozela:2012,Mund:2013,Darius:2017,Markisch:2018}. Other searches include a conversion of a neutron into an anti-neutron~\cite{baldoceolin:1994} or mirror-neutron
%~\cite{Serebrov:2008a}
~\cite{Serebrov:2008a,Altarev:2009}
or a decay into a hypothetical dark matter particle~\cite{Sun:2018}.
\section{Ultra-cold neutrons}
\label{sec-1}
%For two-column wide figures use syntax of figure~\ref{fig-2}

\begin{table}
\centering
\caption{Ultra-cold neutrons can be trapped by the optical potential of matter, by local gravity $g$, or by magnetic fields.}
\small
\label{tab-a}       % Give a unique label
% For LaTeX tables you can use
\begin{tabular}{lll}

\hline
Optical Potential & $\thicksim$100 neV & Material dependencies  \\\hline
Gravity Potential & 100 neV/m & $V$ = $m \times g \times z$ \\\hline
Magnetic Field & 60 neV/T & Zeeman Splitting \\\hline
\end{tabular}
% Or use
%\vspace*{5cm}  % with the correct table height
\end{table}

Neutrons are abundantly produced in a spallation source or a research reactor. At production, these neutrons are very hot with an energy of several MeV. On the other side of the energy scale, some ultra-cold neutron experiments use particles with peV energies in the direction of gravity, about 18 orders of magnitude less. This tremendous reduction is achieved in several steps. In a first step, spallation or fission neutrons thermalize, e.g.,  in a heavy water tank at a temperature of 300 K. The thermal fluxes are distributed in energy close to Maxwellian law. Cold neutrons with energies in the meV range are obtained in a second moderator stage. At the Institut Laue-Langevin this is achieved by a 24 K liquid deuterium cold moderator near the uranium core of the 58.3 MW reactor. As for propagation of light in matter, one can assign a neutron refractive index to materials but it is often less than unity. Thus, one considers the surface of matter as
constituting a potential step of height $V$, see Table 1. Neutrons with
transversal energy $E_{\perp}$ $<$ $V$ will be totally reflected. Following our pragmatic definition, ultra-cold neutrons are, in contrast to faster neutrons, capable of undergoing total external reflection even at normal incidence in a given material or magnetic device~\cite{Steyerl:1977}. When the surface roughness of the mirror is small enough, the ultra-cold neutron reflection is specular. This feature makes it possible to build simple and efficient retroreflectors. Such a neutron mirror makes use of the strong interaction between ultra-cold neutrons and wall nuclei, resulting in an effective spatially extended potential step of the order of 10$^{-15}$ m and 100 neV. We will need a neutron mirror together with gravity for our birthday song in section 3.
\begin{figure*}
\centering
\includegraphics[width=8.cm,clip]{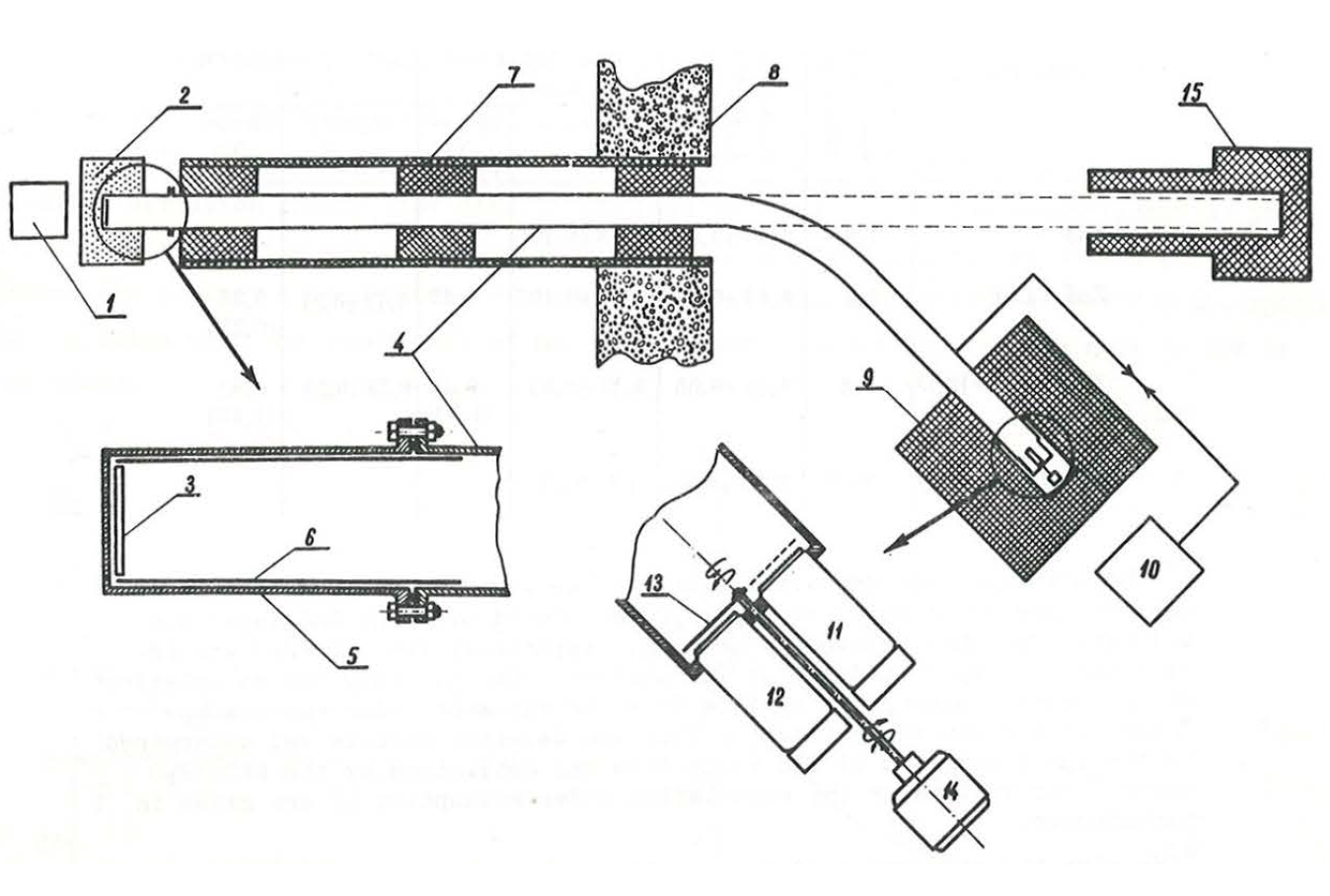}
\includegraphics[width=8.cm,clip]{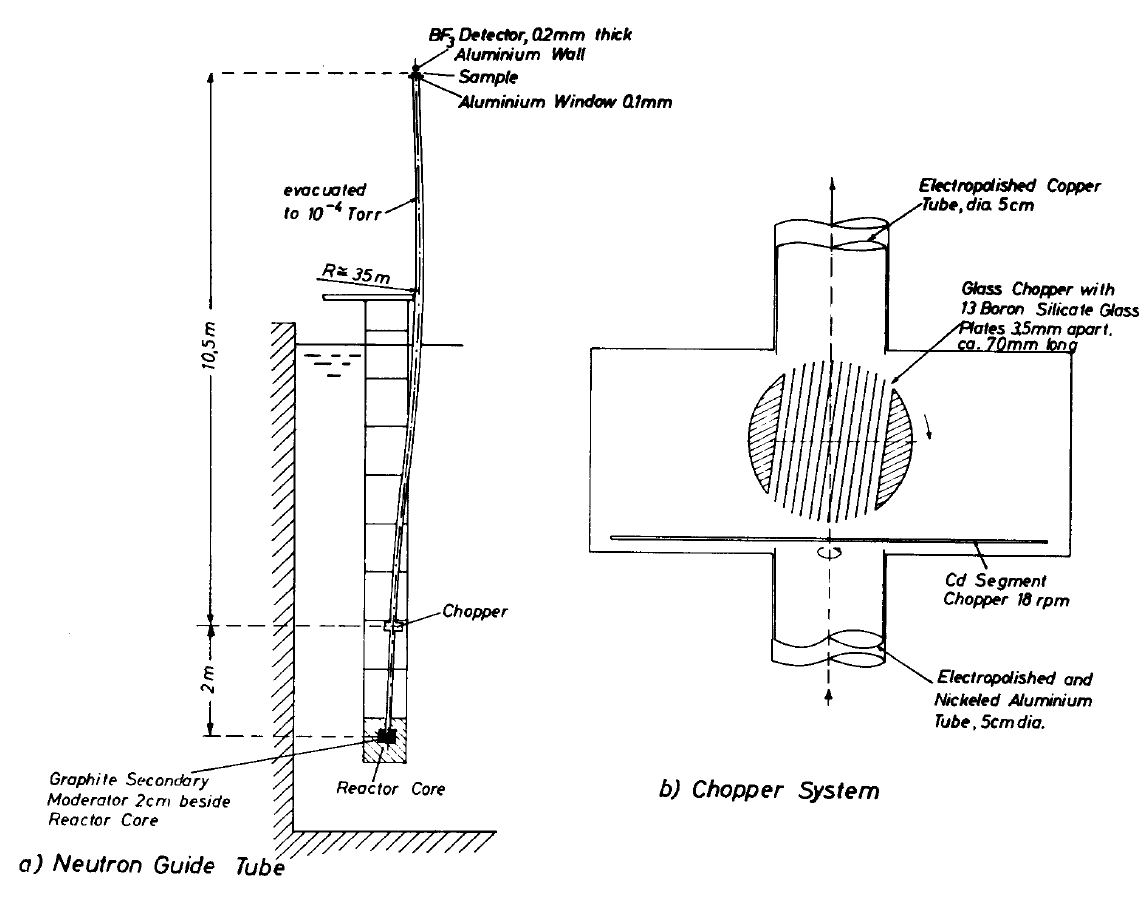}

% Use the relevant command for your figure-insertion program
% to insert the figure file. See example above.
% If not, use
%\vspace*{5cm}       % Give the correct figure height in cm
\caption{The first production of ultra-cold neutrons as an episode during the era of the cold war. Left: experiment of the east, the Dubna experiment by V. I. Lushchikov et al.~\cite{Lushchikov:1969}. The neutron source was the IBR pulsed
reactor operating at an average power of 6 kW and with a flash repetition frequency of one every 5 sec. The only neutrons that could reach the detector in the bent tube in practice were those emitted from moderator labeled 3 with velocities lower than 5.7 m/s for copper~\cite{Lushchikov:1969}. Right: experiment of the west, the Munich experiment by A. Steyerl~\cite{Steyerl:1969}. Very cold neutrons were selected from the
spectrum of a graphite secondary moderator
near the core of the FRM reactor through a vertical neutron
guide tube and energy-analysed with the time-of-flight method. Total cross sections for gold and aluminium
at 300 K were measured for neutron velocities down to 5 m/s. }
\label{fig-2}       % Give a unique label
\end{figure*}

First attempts to produce ultra-cold neutrons started by V. I. Lushchikov, Y. N. Pokotilovskii, A. V. Strelkov, and F. L. Shapiro at Joint Institute for Nuclear Research in  Dubna~\cite{Lushchikov:1969}, and by A. Steyerl at Technische Hochschule M{\"u}nchen~\cite{Steyerl:1969}. Their efforts made first ultra-cold neutrons available exactly 50~years ago. The photo in Fig. 1 shows two of the pioneers during a coffee break at the PPNS workshop.
 Original design drawings together with short descriptions of the experimental arrangements are shown in Fig.~\ref{fig-2}.  In the following years, ultra-cold neutron densities have been drastically increased using more powerful reactor and cold neutron sources at the Petersburg Nuclear Physics Institute (PNPI) and at ILL. Ultra-cold neutrons are taken from the low energy tail of the continuous cold spectrum. At ILL, they are transported vertically upwards by a curved guide, which transmits neutrons below a threshold energy, acting as a low-velocity filter. Neutrons with a velocity of up
to 50~m/s arrive at a rotating turbine with blades consisting of high-quality Ni mirrors. Colliding with the blades moving in the same direction with half their velocity, the neutrons lose almost all their energy and leave the turbine as ultra-cold neutrons with typical velocities of a few meters per second.  They are then guided to several experimental areas. The total output per beam port today is close to 10$^6$~neutrons per second.
%~\cite{Jenke:2018b}.
This instrument, nowadays called PF2, was designed by A. Steyerl and P. Ageron and colleagues and constructed in 1985 \cite{Steyerl:1986}, replacing the former PN5 on level C of the ILL reactor. To date, most experiments using ultra-cold neutrons are limited by counting statistics. Therefore, major efforts are undertaken in order to improve existing ultra-cold neutron sources or develop new concepts.
From the experimental point of view, ideas exist for techniques that could further increase count rates due to a higher ultra-cold neutron phase space density.  One way employs down-scattering of cold neutrons in superfluid $^4$He below 1 K~\cite{Golub:1977,Zimmer:2007,Piegsa:2014}. An alternative approach is the use of solid
deuterium at about 5 K~\cite{Serebrov:1995,Pattie:2016,Anghel:2009}.
Ultra-cold neutron sources are in operation or under construction at many sites, including facilities at the ILL (France), Paul-Scherrer-Institute (Switzerland), University of Mainz (Germany), Los Alamos National Lab (United States), Forschungsneutronenquelle Heinz Maier-Leibnitz (Germany), TRIUMF (Canada), PNPI (Russia), KEK (Japan), North Carolina State University (United States). A comparison of operating ultra-cold neutron sources for fundamental physics measurements can be found in~\cite{Bison:2017}.

%$|7\rangle $$\rightarrow$ $|8\rangle$
%
%$|6\rangle $$\rightarrow$ $|8\rangle$
%
%$|5\rangle $$\rightarrow$ $|8\rangle$
%
%$|4\rangle $$\rightarrow$ $|8\rangle$
%
%$|3\rangle $$\rightarrow$ $|8\rangle$
%
%$|2\rangle $$\rightarrow$ $|8\rangle$
%
%$|1\rangle $$\rightarrow$ $|8\rangle$

\section{Birthday Song for Ultra-Cold Neutrons}
In the following, we present a birthday song in a novel manner for the 50th anniversary of the first production of ultra-cold neutrons. We make use of a musical intonation system that bears no relation to basic notation and basic music theory as applied and used elsewhere. Instead it is based on natural constants and local gravity and addresses two fundamental problems of music theory.

\begin{figure}[h]
% Use the relevant command for your figure-insertion program
% to insert the figure file.
\centering
\includegraphics[width=5.5cm,clip]{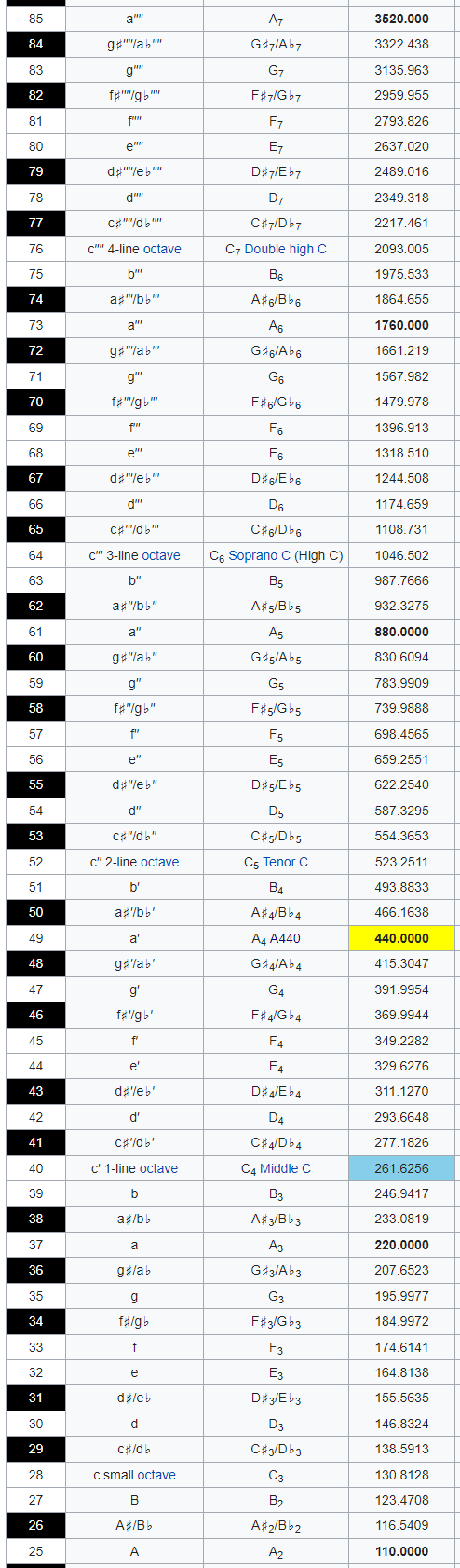}
\caption{Relationship between musical notations and physical frequencies for five octaves between A2 and A7. 1st column: key number on a keyboard, 2nd column: Helmholtz's notation. 3rd column: English notation. Right: corresponding frequencies in equal-tempered intonation. (Figure source: Wikipedia).}
\label{key_frequencies}       % Give a unique label
\end{figure}
\subsection{The Problem of reference for the concert pitch} The first problem is the practical adjustment of absolute pitch of musical instruments. The concert pitch is the reference to which a group of musical instruments is tuned for a performance. Today 440 Hertz is the frequency of the standard tuning tone, called A4 according to Fig.~\ref{key_frequencies}.

Between 1700 and 1820 the concert pitch was pretty constant~\cite{Riemann:1978}: Bach’s tuning fork vibrated at 415.5 Hz and Händel used 422.5 Hz, Berlin was at 422 Hz in 1752 and Mozart’s tuning fork gave 421.6 Hz, the pitch in Paris was 423 Hz. Striving for high brilliance in orchestral sound, a rise in frequency started. In 1858 the following pitch frequencies were used: Paris 449 Hz, Milano 451 Hz, Berlin 452 Hz, London 453 Hz. In 1880 Steinway tuned the pianos to 458 Hz. In an attempt to halt such a pitch inflation the Wien Conference of 1885 recommended 435 Hz as the standard pitch, but in the following decades the pitch rose again to about 443 Hz on average. In 1939, the London Conference of International Federation of the National Standardizing Associations (ISA) recommended that the A be tuned to 440 Hz, now known as concert pitch, confirmed by the International Organization for Standardization as ISO 16 in 1975. Such case that a standard reference was lacking a deeper justification or was subject to fluctuations was also observed in physical science and metrology.

\begin{figure}
% Use the relevant command for your figure-insertion program
% to insert the figure file.
\centering
\sidecaption
\includegraphics[width=7cm,clip]{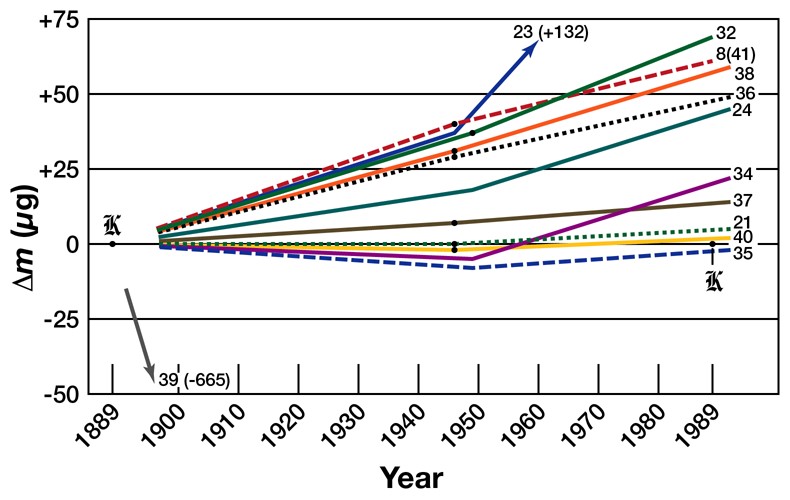}
\caption{Kilogram definition and mass drift over time of national prototypes K21–K40, plus two of the IPK's sister copies: K32 and K8(41) (Wikipedia).}
\label{fig-3}       % Give a unique label
\end{figure}

As an example, we point to the definition of the kilogram as for long established by International System of Units (SI). The standard was based on a   platinum alloy cylinder, the International Prototype Kilogram (IPK), manufactured in 1889, and saved in Saint-Cloud. IPK has diverged from its replicas by approximately 50 $\mu$g since their production, see Fig.~\ref{fig-3}. For this reason it has been desirable to replace the kilogram artifact with a definition based directly on physical constants. Such a definition was approved by the General Conference on Weights and Measures (CGPM) on 16 November 2018.  It is based on constants of nature, in particular the Planck constant, which is now defined, thereby fixing the value of the kilogram in terms of the second and the metre, and thus eliminating the need for the IPK.
%The Planck constant relates a particle’s energy $E$, and hence mass by the Einstein equation $E = mc^2$, to its frequency by $E = \hbar \omega.$

For problems of drifting standards a solution similar to the definition of physical units can be adopted and transfered to musicology. We suggest as a standard for the concert pitch the
resonance frequency of a $|5\rangle \leftrightarrow |8\rangle$ level transition of ultra-cold neutrons in the gravitational field as explained in Section~\ref{sectionqB}. This frequency is based on natural constants as well as the local gravity $g$.

\subsection{The Problem of intonation}

The second problem is related to the choice of intervals between notes. While the basic grid of notes, as represented by the modern piano scale, has been quite uniform over different times and cultures, the fine tuning of those notes has always been a subject of discussion, and has led to the creation of various musical tuning systems. Extreme systems are the modern ``equal-tempered'' intonation, offering a maximum flexibility for switching between tonalities, as opposed to ``just'' or ``pure'' intonation, where the intervals are based on whole-number ratios of frequencies, offering a maximum of clarity within a single tonality. An example is the Pythagorean intonation system, using ratios of 2 and 3 as well as their powers. A property of Pythagorean tuning is an excess of 12 perfect fifths over 7 octaves, which is $(3/2)^{12}$ : $(2/1)^7$ = 531441/524288. The discrepancy is 23.45\% of a semi-tone, or nearly a quarter of a semi tone, and the consequences are a misfit in frequency between enharmonic equal notes like A$_\flat$ and G$_\sharp$. Medieval compositions  avoid such mistunes by a restriction to nine notes, for example B$_{\flat}$-F-C–G–D–A–E–B–F$_{\sharp}$, see for details~\cite{Riemann:1978}.

\begin{figure}[h]
% Use the relevant command for your figure-insertion program
% to insert the figure file.
\centering
\includegraphics[width=5cm,clip]{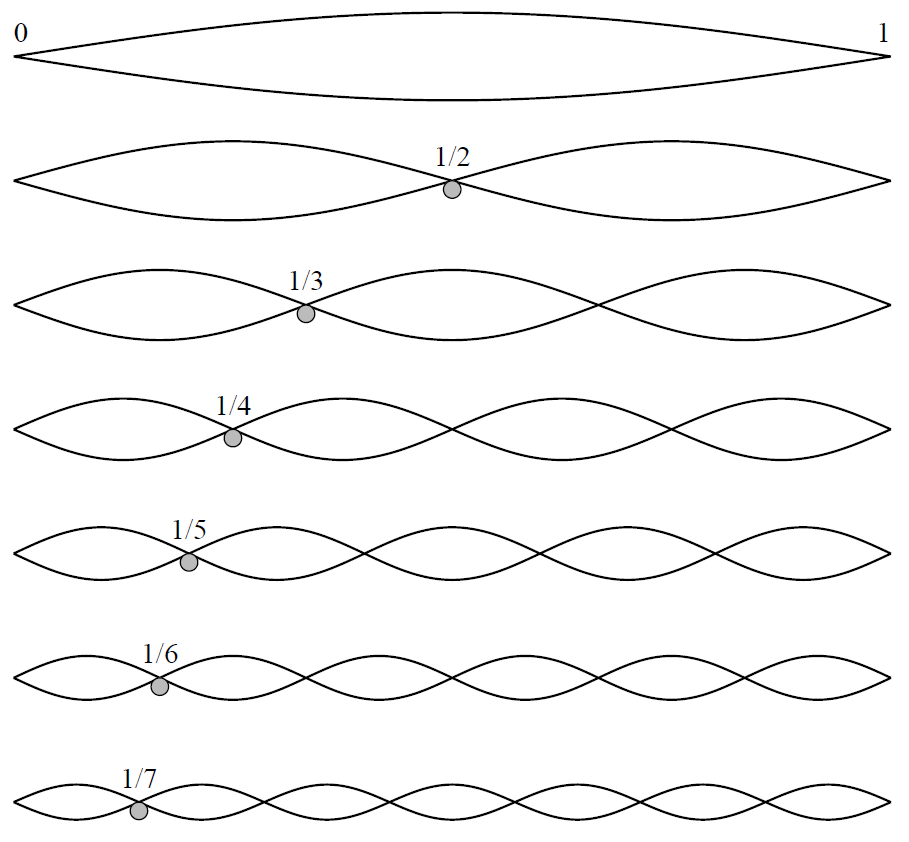}
\caption{Harmonics of a string, based on sine and cosine functions (Wikipedia).}
\label{harmonic}       % Give a unique label
\end{figure}

The number ratios in just intonation are given by the harmonic overtone spectrum of
pitched musical instruments, which are often based
on an acoustic resonator such as a string or a column
of air. The eigenmodes of such resonators form standing
waves, see Fig.~\ref{harmonic}, which correspond to integral frequency ratios.
Numerous such modes oscillate simultaneously.

 The musical pitch of a note is usually perceived as the lowest, fundamental frequency, which may be the one created by vibration over the full length of the string. The musical timbre of the tone is strongly affected by the relative strength of each harmonic. It is interesting to note that the American composer Ben Johnston is experimenting with pure intonation and has proposed the term "extended just intonation" for composition involving ratios that contain prime numbers beyond five (7, 11, 13 etc.), as mentioned at Wikipedia.

 In the equal-tempered intonation system, the modes deviate slightly from the Pythagorean frequency ratios. It parts the octave in 12 equal semitones separated by factors $2^{(1/12)}$. Five octaves of music notes are shown in Fig.~\ref{key_frequencies} in the notation of Helmholtz, in English notation, together with corresponding frequencies in equal-tempered intonation. We introduce an intonation system, which is neither based on the Pythagorean harmonic system nor on the 12$^{th}$ square root of 2.
\begin{figure}[h]
% Use the relevant command for your figure-insertion program
% to insert the figure file.
\centering
\includegraphics[width=8.5cm,clip]{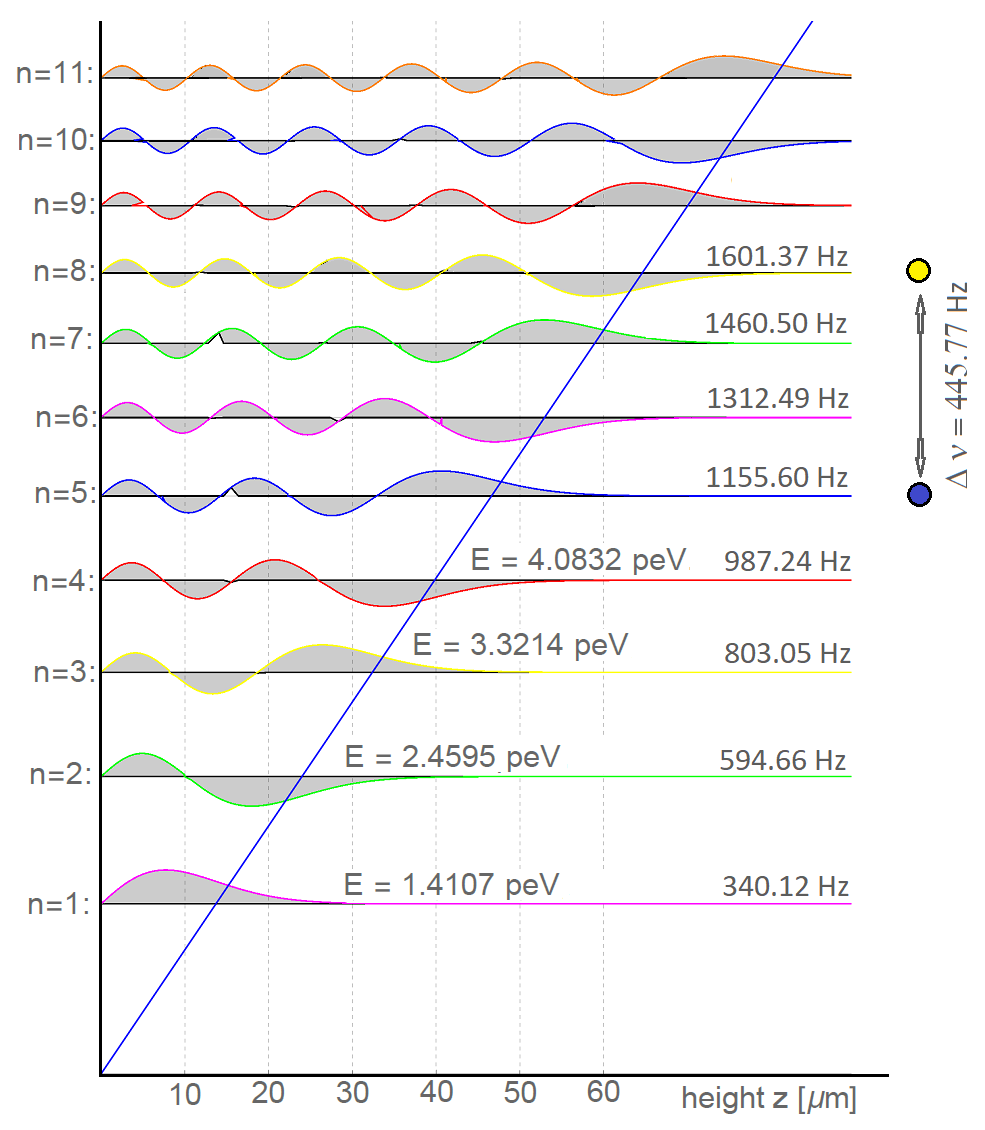}
\caption{Energy eigenvalues and eigenfunctions of a neutron bound in the gravity potential of the Earth and corresponding frequency equivalents. By oscillating the mirror with a frequency, which correspond to the energy difference between quantum states, transitions are introduced. The oscillation frequencies are in the acoustic frequency range.}
\label{qBounce}       % Give a unique label
\end{figure}

\subsection{Neutron's intonation \\ and the quantum bouncing ball}\label{sectionqB}We approach both mentioned problems of music theory by resorting to the natural frequency spectrum given by an ultra-cold neutron prepared as quantum bouncing ball~\cite{Abele_2012}.
The neutron is bound between a flat lying mirror and the raising potential of gravity above, $V=m g z$,
where $m$ denotes the neutron mass, $g$ the local gravity, and $z$ the distance above the mirror.
Every bound system in quantum mechanics has discrete energy levels. In our case the lowest energy
eigenvalues $E_n$ ($n$ = 1, 2, 3, 4, 5) are 1.4107 peV, 2.4595 peV, 3.3214 peV, 4.0832 peV,
and 4.77906 peV. In Fig.~\ref{qBounce} they are shown together with the corresponding neutron wavefunctions, the well-known Airy functions. Similarly, in the neutron whispering gallery quantum states occur in a binding well formed by the centrifugal potential and a circular boundary~\cite{Nesvizhevsky:2010}.

The interesting point is that one can drive transitions between these eigenstates
by vibrating the mirror. Within the qBounce experiment~\cite{Abele:2010,Jenke:2011,Jenke:2014,Jenke:2018,Cronenberg:2018}, several resonance spectroscopy measurements with
different geometric parameters have been performed, resulting in different resonance frequencies and widths.
In general, the oscillator frequency at resonance for a transition between states with energies $E_p$ and $E_q$ is
\begin{equation}
\nu_{p,q}=\frac{E_q-E_p}{h}=\nu_q-\nu_p.
\label{eq-1}\end{equation}
The transfer from state $|p\rangle$ to $|q\rangle$ is referred to as Rabi transition, which can be induced by applying the right frequency
in the acoustic range. For example, the  $|1\rangle \leftrightarrow |2\rangle$ transition corresponds
to the frequency $\nu_{1,2}$ = 254.542 Hz, the transition $|1\rangle \leftrightarrow |3\rangle$ has a frequency
of $\nu_{1,3}$ = 462.94 Hz, and $\nu_{5,8}$ = 445.77 Hz is close to concert pitch of 444 Hz,
which is used by many orchestras although the official value is 440 Hz.

\begin{table}
  \centering
  \begin{tabular}{ccc}
  Location & Local  & Concert pitch [Hz] \\
  & acceleration $g$ & based on  \\
  &[m/s$^2$]& ultra-cold neutrons\\ \hline
       Amsterdam & 9.813 & 446.01 \\
       Athens & 9.800 & 445.61 \\
         Bangkok & 9.783 &  445.10\\
          Cape Town & 9.796 & 445.49 \\
           Chicago &9.803  & 445.70 \\
           Grenoble &9.805&445.77\\
           Helsinki & 9.819 & 446.19 \\
           Havana & 9.788& 445.25 \\
           Istanbul & 9.808 & 445.85 \\
           Jakarta & 9.781 &445.04  \\
           London & 9.796 & 445.49 \\
           Mexico City & 9.779 & 444.97 \\
           Paris & 9.809 &  445.88\\
           San Francisco & 9.800 & 445.61 \\
       Sydney & 9.797  & 445.52 \\
       Vienna & 9.808 & 445.85 \\
       Zurich & 9.807 & 445.82 \\
  \end{tabular}
  \caption{A selection of cities with corresponding local acceleration $g$. Each city has its own character. To take this fact into account it is suggested to establish a local concert pitch for music performances based on local $g$ and on ultra-cold neutrons.}
  \label{Amsterdam}
\end{table}

\begin{figure}[h]
% Use the relevant command for your figure-insertion program
% to insert the figure file.
\centering
\includegraphics[width=8cm,clip]{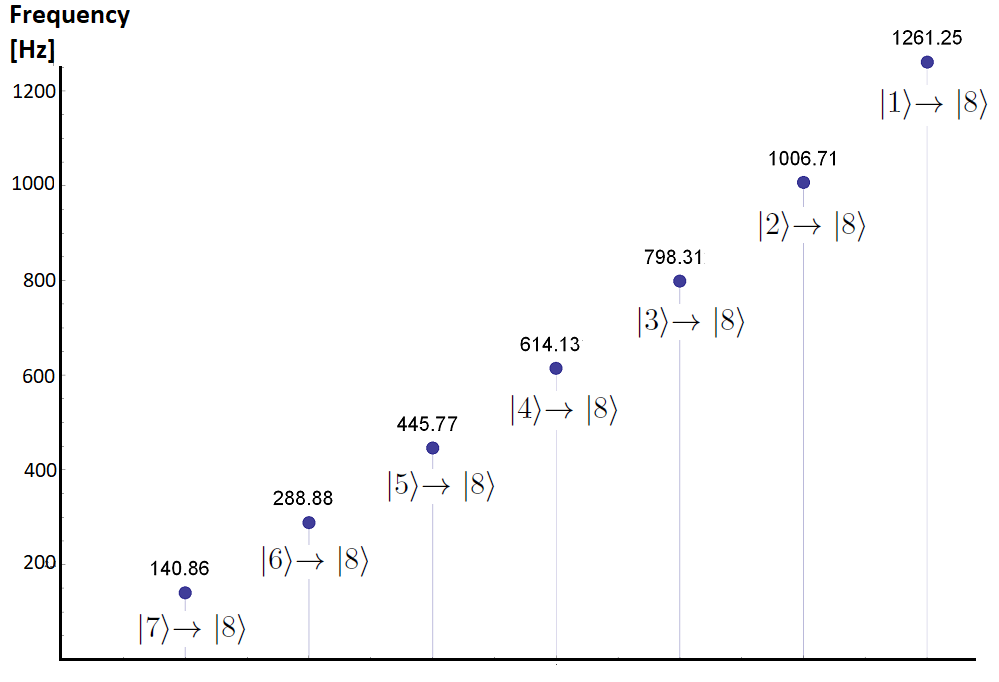}
\caption{Planck's constant $h$, the mass of the neutron $m$ and local gravity $g$ define a series of frequency, here exemplarily demonstrated for acoustic transitions between  $|7\rangle $$\rightarrow$ $|8\rangle$ at 140.86 Hz and $|1\rangle$  $\rightarrow$ $|8\rangle$ at 1261.25 Hz. In this paper the frequencies $\nu$ are rounded to two decimal places. Such a frequency resolution is reached during a time span of 10 s taking the classical uncertainty relationship $\Delta \omega$ $\times$ $\Delta t$ = $\frac{1}{2}$ into consideration.}
%in agreement with a reachable by a time span of 10 s, motivated by the dictated by the.}
\label{18}       % Give a unique label
\end{figure}

We therefore suggest to use the acoustic $|5\rangle \leftrightarrow |8\rangle$ transition as a definition
for the concert pitch, satisfying our requirement for natural constants, the mass of the neutron $m$ and Planck's constant $h$, together with the
local gravity $g$.
\begin{equation}
\nu_{5,8}=\frac{E_8-E_5}{h}=\nu_8-\nu_5 = 445.77~\mathrm{Hz}.
\end{equation}
We would like to note that, as a particularity of a quantum system involving gravity as a binding force, this definition is still open to display regional diversity. This, however, is not due to an arbitrary choice, as for the pitch in a concert hall, but defined by nature. A local gravity $g$ = 9.797 m/s$^2$ at Sydney
results in a concert pitch of 445.52 Hz, whereas the opera house in Helsinki with  $g$ = 9.819 would
use a pitch of 446.19 Hz. Concert pitches at other locations can be found in Table~\ref{Amsterdam}.
For our calculations we use the local gravity $g$ = 9.80507 m/s$^2$ at our experimental setup PF2 in Grenoble.

%This tuning allows us to present a birthday song for ultra-cold neutrons based on their own features.
%The acoustic transitions between $|7\rangle $$\rightarrow$ $|8\rangle$  at $140.835 ~\mathrm{Hz}$ and $|1\rangle$  $\rightarrow$ $|8\rangle$ at 1263.25 Hz define a series of frequency, see Fig.~\ref{18}, which is sometimes close to the piano key notes of Fig.~\ref{key_frequencies}. A more complete set of frequencies can be found in Table~\ref{tab-1}, where neutron's intonation system is represented as a 30 $\times$ 14 matrix $M$ with $q$ corresponding to the $q^{th}$ energy eigenstate, and the matrix element $M_{pq}$ displays the transition frequency $\nu_{pq}$ of Equation~\ref{eq-1}. Shown is the frequency range for transitions between $|14\rangle$ $ \leftrightarrow$ $|15\rangle$ at 112.45 Hz and  $|1\rangle$ $\leftrightarrow $ $|30\rangle$ at 3585.57 Hz spanning the piano key note frequencies from A2 to A7 in Fig.~\ref{key_frequencies}.
\begin{figure*}[h]
\centering
\includegraphics[width=17cm,clip]{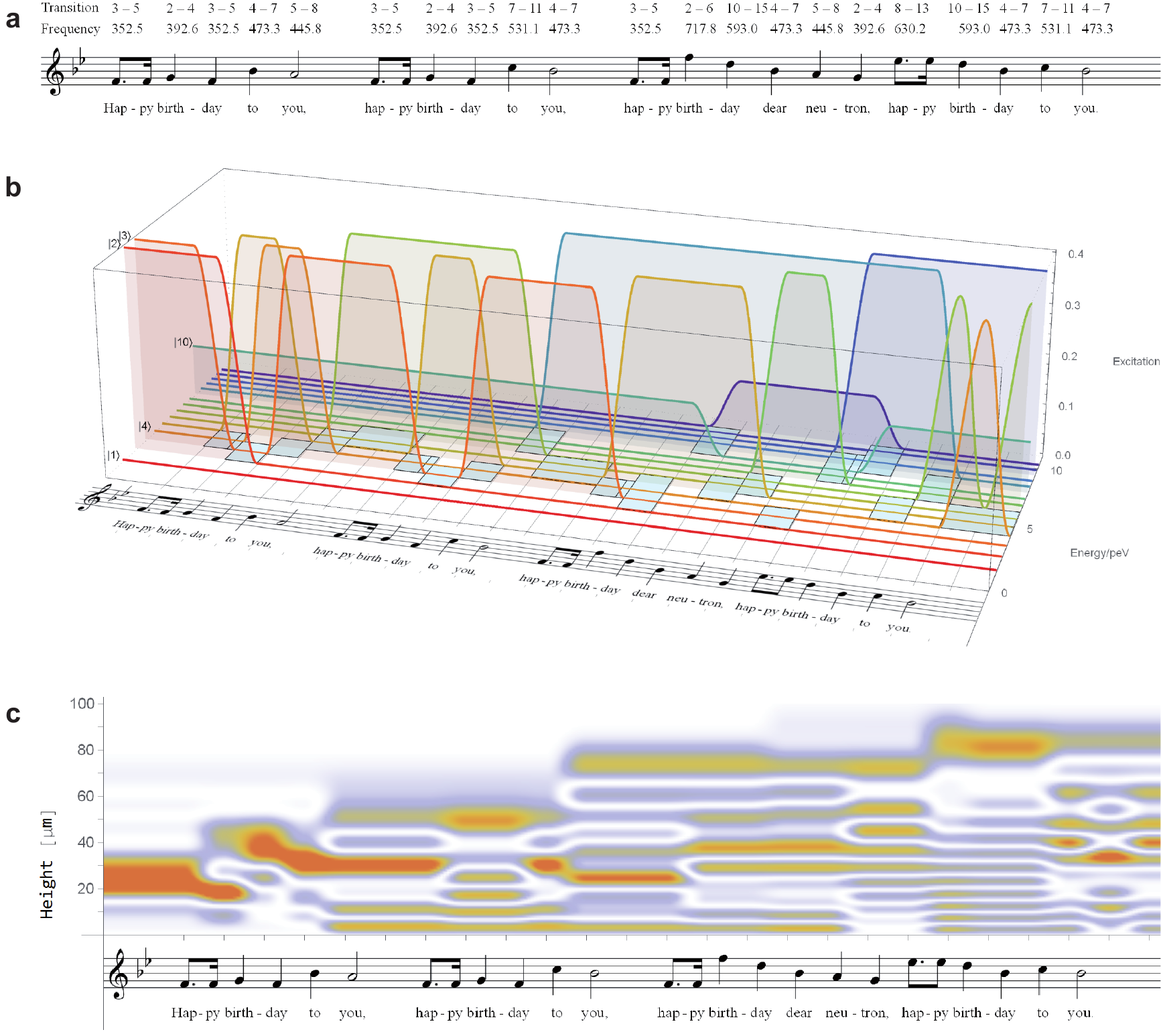}
\caption{A birthday song for ultra-cold neutrons
	(a) Level couplings and frequencies used for the birthday song
	(b) Each song note couples two levels, indicated by the light blue rectangles. As long as the note is sounding the level amplitudes make a Rabi oscillation. The tone duration of a quarter note creates a $\pi$ flip.
	(c) Neutron wave function $|\psi|^2$ of the dancing neutron.
}
\label{song}       % Give a unique label
\end{figure*}
The transitions between $|7\rangle $$\rightarrow$ $|8\rangle$ at $140.835 ~\mathrm{Hz}$ and $|1\rangle$  $\rightarrow$ $|8\rangle$ at 1263.25 Hz define a series of frequency, see Fig.~\ref{18}. Taking all transitions  $|p\rangle $$\rightarrow$ $|q\rangle$ into account one obtains neutron's intonation system, here shown in Table~\ref{tab-1}~in a matrix of frequencies up to level 30. Many frequencies are close to notes of the modern piano scale, cf. Fig.~\ref{key_frequencies}, and we therefore suggest to consider the neutron transition frequencies a musical instrument. Selected frequencies can be used to tune the piano scale, addressing the second problem mentioned above. However, the complete spectrum is much more diverse than a standard musical instrument. Each pair of levels has a unique transition frequency, as a consequence of the Airy functions.
In principle the spectrum offers vast musical possibilities but for now
we restrict ourselves to selected frequencies suitable for the traditional song "Happy Birthday".
We play the song  in B$\flat$ major starting with an f. With 440 Hz tuning and equal temperament
the f corresponds to 349.2282 Hz. In the neutron tuning we use the $|3\rangle$  $\rightarrow$ $|5\rangle$ transition corresponding to 352.548 Hz and continue with other transitions as shown in Fig.~\ref{song}a. The audio file can be downloaded from our web site.

\begin{table*}[h]
\centering
\tiny
\begin{tabular}{lcccccccccccccc}
&$|1\rangle$&$|2\rangle$&$|3\rangle$&$|4\rangle$&$|5\rangle$&$|6\rangle$&$|7\rangle$&$|8\rangle$&$|9\rangle$&$|10\rangle$&$|11\rangle$&$|12\rangle$&$|13\rangle$&$|14\rangle$\\
$|1\rangle$& 0. & 254.542 & 462.939 & 647.12 & 815.486 & 972.374 & 1120.39 & 1261.25 & 1396.17 & 1526.04 & 1651.53 & 1773.19 & 1891.44 & 2006.64 \\
$|2\rangle$& 254.542 & 0. & 208.396 & 392.578 & 560.944 & 717.831 & 865.846 & 1006.71 & 1141.63 & 1271.49 & 1396.99 & 1518.65 & 1636.9 & 1752.1 \\
$|3\rangle$& 462.939 & 208.396 & 0. & 184.182 & 352.548 & 509.435 & 657.45 & 798.312 & 933.231 & 1063.1 & 1188.59 & 1310.25 & 1428.5 & 1543.7 \\
$|4\rangle$& 647.12 & 392.578 & 184.182 & 0. & 168.366 & 325.254 & 473.269 & 614.131 & 749.049 & 878.915 & 1004.41 & 1126.07 & 1244.32 & 1359.52 \\
$|5\rangle$& 815.486 & 560.944 & 352.548 & 168.366 & 0. & 156.888 & 304.903 & 445.765 & 580.683 & 710.549 & 836.045 & 957.704 & 1075.96 & 1191.16 \\
$|6\rangle$& 972.374 & 717.831 & 509.435 & 325.254 & 156.888 & 0. & 148.015 & 288.877 & 423.795 & 553.662 & 679.157 & 800.816 & 919.069 & 1034.27 \\
$|7\rangle$& 1120.39 & 865.846 & 657.45 & 473.269 & 304.903 & 148.015 & 0. & 140.862 & 275.78 & 405.647 & 531.142 & 652.801 & 771.054 & 886.253 \\
$|8\rangle$& 1261.25 & 1006.71 & 798.312 & 614.131 & 445.765 & 288.877 & 140.862 & 0. & 134.918 & 264.785 & 390.28 & 511.939 & 630.192 & 745.391 \\
$|9\rangle$& 1396.17 & 1141.63 & 933.231 & 749.049 & 580.683 & 423.795 & 275.78 & 134.918 & 0. & 129.866 & 255.362 & 377.021 & 495.274 & 610.473 \\
$|10\rangle$& 1526.04 & 1271.49 & 1063.1 & 878.915 & 710.549 & 553.662 & 405.647 & 264.785 & 129.866 & 0. & 125.495 & 247.154 & 365.407 & 480.606 \\
$|11\rangle$& 1651.53 & 1396.99 & 1188.59 & 1004.41 & 836.045 & 679.157 & 531.142 & 390.28 & 255.362 & 125.495 & 0. & 121.659 & 239.912 & 355.111 \\
$|12\rangle$& 1773.19 & 1518.65 & 1310.25 & 1126.07 & 957.704 & 800.816 & 652.801 & 511.939 & 377.021 & 247.154 & 121.659 & 0. & 118.253 & 233.452 \\
$|13\rangle$& 1891.44 & 1636.9 & 1428.5 & 1244.32 & 1075.96 & 919.069 & 771.054 & 630.192 & 495.274 & 365.407 & 239.912 & 118.253 & 0. & 115.199 \\
$|14\rangle$& 2006.64 & 1752.1 & 1543.7 & 1359.52 & 1191.16 & 1034.27 & 886.253 & 745.391 & 610.473 & 480.606 & 355.111 & 233.452 & 115.199 & 0. \\
$|15\rangle$& 2119.08 & 1864.54 & 1656.14 & 1471.96 & 1303.59 & 1146.71 & 998.691 & 857.829 & 722.91 & 593.044 & 467.549 & 345.89 & 227.637 & 112.438 \\
$|16\rangle$& 2229. & 1974.46 & 1766.06 & 1581.88 & 1413.52 & 1256.63 & 1108.61 & 967.753 & 832.834 & 702.968 & 577.472 & 455.813 & 337.56 & 222.361 \\
$|17\rangle$& 2336.62 & 2082.08 & 1873.69 & 1689.5 & 1521.14 & 1364.25 & 1216.24 & 1075.37 & 940.455 & 810.588 & 685.093 & 563.434 & 445.181 & 329.982 \\
$|18\rangle$& 2442.12 & 2187.58 & 1979.18 & 1795. & 1626.64 & 1469.75 & 1321.73 & 1180.87 & 1045.95 & 916.087 & 790.592 & 668.933 & 550.68 & 435.481 \\
$|19\rangle$& 2545.66 & 2291.12 & 2082.72 & 1898.54 & 1730.17 & 1573.28 & 1425.27 & 1284.41 & 1149.49 & 1019.62 & 894.127 & 772.468 & 654.215 & 539.016 \\
$|20\rangle$& 2647.37 & 2392.83 & 2184.43 & 2000.25 & 1831.88 & 1674.99 & 1526.98 & 1386.12 & 1251.2 & 1121.33 & 995.837 & 874.178 & 755.925 & 640.726 \\
$|21\rangle$& 2747.38 & 2492.83 & 2284.44 & 2100.26 & 1931.89 & 1775. & 1626.99 & 1486.12 & 1351.21 & 1221.34 & 1095.84 & 974.186 & 855.933 & 740.734 \\
$|22\rangle$& 2845.79 & 2591.25 & 2382.85 & 2198.67 & 2030.3 & 1873.42 & 1725.4 & 1584.54 & 1449.62 & 1319.75 & 1194.26 & 1072.6 & 954.346 & 839.147 \\
$|23\rangle$& 2942.7 & 2688.16 & 2479.77 & 2295.58 & 2127.22 & 1970.33 & 1822.32 & 1681.45 & 1546.54 & 1416.67 & 1291.17 & 1169.51 & 1051.26 & 936.063 \\
$|24\rangle$& 3038.21 & 2783.67 & 2575.27 & 2391.09 & 2222.72 & 2065.84 & 1917.82 & 1776.96 & 1642.04 & 1512.18 & 1386.68 & 1265.02 & 1146.77 & 1031.57 \\
$|25\rangle$& 3132.39 & 2877.84 & 2669.45 & 2485.27 & 2316.9 & 2160.01 & 2012. & 1871.13 & 1736.22 & 1606.35 & 1480.85 & 1359.2 & 1240.94 & 1125.74 \\
$|26\rangle$& 3225.3 & 2970.76 & 2762.36 & 2578.18 & 2409.81 & 2252.93 & 2104.91 & 1964.05 & 1829.13 & 1699.26 & 1573.77 & 1452.11 & 1333.86 & 1218.66 \\
$|27\rangle$& 3317.02 & 3062.48 & 2854.08 & 2669.9 & 2501.53 & 2344.65 & 2196.63 & 2055.77 & 1920.85 & 1790.98 & 1665.49 & 1543.83 & 1425.58 & 1310.38 \\
$|28\rangle$& 3407.6 & 3153.06 & 2944.66 & 2760.48 & 2592.12 & 2435.23 & 2287.21 & 2146.35 & 2011.43 & 1881.57 & 1756.07 & 1634.41 & 1516.16 & 1400.96 \\
$|29\rangle$& 3497.1 & 3242.56 & 3034.17 & 2849.98 & 2681.62 & 2524.73 & 2376.71 & 2235.85 & 2100.93 & 1971.07 & 1845.57 & 1723.91 & 1605.66 & 1490.46 \\
$|30\rangle$& 3585.57 & 3331.03 & 3122.63 & 2938.45 & 2770.09 & 2613.2 & 2465.18 & 2324.32 & 2189.4 & 2059.54 & 1934.04 & 1812.38 & 1694.13 & 1578.93 \\
\end{tabular}
\caption{
Neutron's intonation system based on accoustic resonant transitions~\cite{Cronenberg:2018}, here in a 30 $\times$ 14 matrix $M$. Shown is the frequency range for transitions between $|14\rangle$ $ \leftrightarrow$ $|15\rangle$ at 112.45 Hz and  $|1\rangle$ $\leftrightarrow $ $|30\rangle$ at 3585.57 Hz spanning the piano key notes from A2 to A7 in Fig~\ref{key_frequencies}. We can consider this matrix as an instrument, which is able to perform pieces of music with these frequencies, and perform "Happy Birthday to you" for the ultra-cold neutron.
The frequencies have been calculated for the local gravity $g$~=~9.80507~m/s$^2$ at our experimental setup PF2 in Grenoble.}
\label{tab-1}       % Giv
%\vspace*{15cm}  % with the correct table height
\end{table*}

As it's their 50 year's birthday, it's not us but the ultra-cold neutrons to listen to the song. It's us to play it and watch their reaction. We send the signal to the neutron mirror and let it vibrate. Each note starts a Rabi transition between the corresponding two levels. The rhythm (tone duration) determines the coupling time and therefore the amount of amplitude change. We adjust the vibration strength such that a quarter note induces an exact $\pi$ flip, swapping the complete amplitude between two levels. As initial state we prepare the superposition $\sqrt{0.4}\, |2\rangle + \sqrt{0.4}\, |3\rangle + \sqrt{0.2}\, |10\rangle $.
Figure~\ref{song}b shows the couplings and the excitation amplitudes of the levels during the song.
Figure~\ref{song}c shows the resulting neutron probability density as a function of time and vertical position over the mirror. The neutron is dancing to its birthday song:\newline
\textbf{Happy Birthday Ultra-Cold Neutron!}

% = end of block comment

\section*{Ackowledgement}
H.A. would like to thank G. Horvath, who presented the neutron's time evolution with respect to the first 30 s of Wolfgang
Amadeus Mozart’s Opera Cosi fan tutte, Aria Fiordiligi, Come scolgi.
\newpage
\newpage


\begin{thebibliography}{review}
\bibitem{Lushchikov:1969}
V.I. Lushchikov, Y.N. Pokotilovskii, A.V. Strelkov, F.L. Shapiro, JETP Lett.
  (U.S.S.R., Engl. Transl.) \textbf{9}, 40 (1969)

\bibitem{Steyerl:1969}
A.~Steyerl, Phys. Lett. B \textbf{29}, 33 (1969)

\bibitem{Cronenberg:2018}
G.~Cronenberg, P.~Brax, H.~Filter, P.~Geltenbort, T.~Jenke, G.~Pignol,
  M.~Pitschmann, M.~Thalhammer, H.~Abele, Nature Physics \textbf{14}, 1022
  (2018)

\bibitem{Riemann:1978}
Riemann, \emph{Musiklexikon} ({F.A. Brockhaus, B. Schotts Söhne, Wiesbaden,
  Mainz}, 1978)

\bibitem{Steyerl:1977}
A.~Steyerl, \emph{Very low energy neutrons} (Springer Berlin Heidelberg,
  Berlin, Heidelberg, 1977), pp. 57--130, ISBN 978-3-540-37543-2,
  \urlstyle{tt}\url{https://doi.org/10.1007/BFb0041487}

\bibitem{Pendlebury:2015}
J.M. Pendlebury, S.~Afach, N.J. Ayres, C.A. Baker, G.~Ban, G.~Bison, K.~Bodek,
  M.~Burghoff, P.~Geltenbort, K.~Green et~al., Phys. Rev. D \textbf{92}, 092003
  (2015)

\bibitem{Ramsey:1949}
N.F. Ramsey, Phys. Rev. \textbf{76}, 996 (1949)

\bibitem{Altarev:1978se}
I.S. Altarev et~al., JETP Lett. \textbf{29}, 730 (1979)

\bibitem{Altarev:1981}
I.S. Altarev, Y.V. Borisov, N.V. Borovikova, A.B. Brandin, A.I. Egorov, V.F.
  Ezhov, S.N. Ivanov, V.M. Lobashev, V.A. Nazarenko, V.L. Ryabov et~al., Phys.
  Lett. B \textbf{102}, 13 (1981)

\bibitem{Pendlebury:1984}
J.M. Pendlebury, K.F. Smith, R.~Golub, J.~Byrne, T.J.L. McComb, T.J. Sumner,
  S.M. Burnett, A.R. Taylor, B.~Heckel, N.F. Ramsey et~al., Phys. Lett. B
  \textbf{136}, 327 (1984)

\bibitem{Altarev:1986}
I.~Altarev, {et al.}, JETP Lett. (U.S.S.R., Engl. Transl.) \textbf{44}, 460
  (1986)

\bibitem{Smith:1990}
K.F. Smith, N.~Crampin, J.M. Pendlebury, D.J. Richardson, D.~Shiers, K.~Green,
  A.I. Kilvington, J.~Moir, H.B. Prosper, D.~Thompson et~al., Phys. Lett. B
  \textbf{234}, 191 (1990)

\bibitem{Altarev:1992}
I.S. Altarev, Y.V. Borisov, N.V. Borovikova, S.N. Ivanov, E.A. Kolomensky, M.S.
  Lasakov, V.M. Lobashev, V.A. Nazarenko, A.N. Pirozhkov, A.P. Serebrov et~al.,
  Phys. Lett. B \textbf{276}, 242 (1992)

\bibitem{Altarev:1996}
I.S. Altarev, Y.V. Borisov, N.V. Borovikova, A.I. Egorov, S.N. Ivanov, E.A.
  Kolomensky, M.S. Lasakov, V.M. Lobashev, V.A. Nazarenko, A.N. Pirozhkov
  et~al., Phys. Atom. Nuclei \textbf{59}, 1152 (1996)

\bibitem{Harris:1999}
P.G. Harris, C.A. Baker, K.~Green, P.~Iaydjiev, S.~Ivanov, D.J.R. May, J.M.
  Pendlebury, D.~Shiers, K.F. Smith, M.~{van der Grinten} et~al., Phys. Rev.
  Lett. \textbf{82}, 904 (1999)

\bibitem{Baker:2006}
C.A. Baker, D.D. Doyle, P.~Geltenbort, K.~Green, M.G.D. {van der Grinten}, P.G.
  Harris, P.~Iaydjiev, S.N. Ivanov, D.J.R. May, J.M. Pendlebury et~al., Phys.
  Rev. Lett. \textbf{97}, 131801 (2006)

\bibitem{Abel:2017}
C.~Abel, N.J. Ayres, G.~Ban, G.~Bison, K.~Bodek, V.~Bondar, M.~Daum,
  M.~Fairbairn, V.V. Flambaum, P.~Geltenbort et~al., Phys. Rev. X \textbf{7},
  041034 (2017)

\bibitem{Dubbers:2011}
D.~Dubbers, M.G. Schmidt, Rev. Mod. Phys. \textbf{83}, 1111 (2011)

\bibitem{ABELE:2008}
H.~Abele, Prog. Part. Nucl. Phys. \textbf{60}, 1 (2008)

\bibitem{Mampe:1989a}
W.~Mampe, P.~Ageron, C.~Bates, J.M. Pendlebury, A.~Steyerl, Phys. Rev. Lett.
  \textbf{63}, 593 (1989)

\bibitem{Arzumanov:2000}
S.~Arzumanov, L.~Bondarenko, S.~Chernyavsky, W.~Drexel, A.~Fomin,
  P.~Geltenbort, V.~Morozov, Y.~Panin, J.~Pendlebury, K.~Schreckenbach, Phys.
  Lett. B \textbf{483}, 15 (2000)

\bibitem{Serebrov:2008}
A.P. Serebrov, V.E. Varlamov, A.G. Kharitonov, A.K. Fomin, Y.N. Pokotilovski,
  P.~Geltenbort, I.A. Krasnoschekova, M.S. Lasakov, R.R. Taldaev, A.V.
  Vassiljev et~al., Phys. Rev. C \textbf{78}, 035505 (2008)

\bibitem{Ezhov:2009}
V.~Ezhov, A.~Andreev, G.~Ban, B.~Bazarov, P.~Geltenbort, F.~Hartman,
  A.~Glushkov, M.~Groshev, V.~Knyazkov, N.~Kovrizhnykh et~al., Nucl. Instrum.
  Meth. Phys. Res. A \textbf{611}, 167 (2009)

\bibitem{Pichlmaier:2010}
A.~Pichlmaier, V.~Varlamov, K.~Schreckenbach, P.~Geltenbort, Phys. Lett. B
  \textbf{693}, 221 (2010)

\bibitem{Serebrov:2018}
A.P. Serebrov, E.A. Kolomensky, A.K. Fomin, I.A. Krasnoshchekova, A.V.
  Vassiljev, D.M. Prudnikov, I.V. Shoka, A.V. Chechkin, M.E. Chaikovskiy, V.E.
  Varlamov et~al., Phys. Rev. C \textbf{97}, 055503 (2018)

\bibitem{Pattie:2018}
R.W. Pattie, N.B. Callahan, C.~Cude-Woods, E.R. Adamek, L.J. Broussard, S.M.
  Clayton, S.A. Currie, E.B. Dees, X.~Ding, E.M. Engel et~al., Science
  \textbf{360}, 627 (2018)

\bibitem{Ezhov:2018}
V.F. Ezhov, A.Z. Andreev, G.~Ban, B.A. Bazarov, P.~Geltenbort, A.G. Glushkov,
  V.A. Knyazkov, N.A. Kovrizhnykh, G.B. Krygin, O.~{Naviliat-Cuncic} et~al.,
  JETP Lett. \textbf{107}, 671 (2018)

\bibitem{Pattie:2009}
R.W. Pattie, J.~Anaya, H.O. Back, J.G. Boissevain, T.J. Bowles, L.J. Broussard,
  R.~Carr, D.J. Clark, S.~Currie, S.~Du et~al., Phys. Rev. Lett. \textbf{102},
  012301 (2009)

\bibitem{Mendenhall:2013}
M.P. Mendenhall, R.W. Pattie, Y.~Bagdasarova, D.B. Berguno, L.J. Broussard,
  R.~Carr, S.~Currie, X.~Ding, B.W. Filippone, A.~García et~al. (UCNA
  Collaboration), Phys. Rev. C \textbf{87}, 032501 (2013)

\bibitem{Brown:2018}
M.A.P. Brown, E.B. Dees, E.~Adamek, B.~Allgeier, M.~Blatnik, T.J. Bowles, L.J.
  Broussard, R.~Carr, S.~Clayton, C.~{Cude-Woods} et~al. (UCNA Collaboration),
  Phys. Rev. C \textbf{97}, 035505 (2018)

\bibitem{Luschikov:1978}
V.I. Luschikov, A.I. Frank, JETP Lett. \textbf{28}, 559 (1978)

\bibitem{Rauch:2015}
H.~Rauch, S.A. Werner, \emph{Neutron {{Interferometry}}: {{Lessons}} in
  {{Experimental Quantum Mechanics}}, {{Wave}}-{{Particle Duality}}, and
  {{Entanglement}}} ({Oxford University Press}, 2015), ISBN 978-0-19-871251-0

\bibitem{Frank:2006}
A.I. Frank, P.~Geltenbort, G.V. Kulin, D.V. Kustov, V.G. Nosov, A.N. Strepetov,
  JETP Lett. \textbf{84}, 363 (2006)

\bibitem{Frank:2011}
A.I. Frank, P.~Geltenbort, M.~Jentschel, D.V. Kustov, G.V. Kulin, A.N.
  Strepetov, JETP Lett. \textbf{93}, 361 (2011)

\bibitem{Durstberger-Rennhofer:2011}
K.~{Durstberger-Rennhofer}, T.~Jenke, H.~Abele, Phys. Rev. D \textbf{84},
  036004 (2011)

\bibitem{Byrne:1996}
J.~Byrne, P.G. Dawber, C.G. Habeck, S.J. Smidt, J.A. Spain, A.P. Williams,
  Europhysics Letters ({EPL}) \textbf{33}, 187 (1996)

\bibitem{Dewey:2003}
M.S. Dewey, D.M. Gilliam, J.S. Nico, F.E. Wietfeldt, X.~Fei, W.M. Snow, G.L.
  Greene, J.~Pauwels, R.~Eykens, A.~Lamberty et~al., Phys. Rev. Lett.
  \textbf{91}, 152302 (2003)

\bibitem{Nico:2005}
J.S. Nico, M.S. Dewey, D.M. Gilliam, F.E. Wietfeldt, X.~Fei, W.M. Snow, G.L.
  Greene, J.~Pauwels, R.~Eykens, A.~Lamberty et~al., Phys. Rev. C \textbf{71},
  055502 (2005)

\bibitem{Yue:2013}
A.T. Yue, M.S. Dewey, D.M. Gilliam, G.L. Greene, A.B. Laptev, J.S. Nico, W.M.
  Snow, F.E. Wietfeldt, Phys. Rev. Lett. \textbf{111}, 222501 (2013)

\bibitem{Wietfeldt:2011}
F.E. Wietfeldt, G.L. Greene, Rev. Mod. Phys. \textbf{83}, 1173 (2011)

\bibitem{Bopp:1986}
P.~Bopp, D.~Dubbers, L.~Hornig, E.~Klemt, J.~Last, H.~Schütze, S.J. Freedman,
  O.~Schärpf, Phys. Rev. Lett. \textbf{56}, 919 (1986)

\bibitem{Schreckenbach:1995}
K.~Schreckenbach, P.~Liaud, R.~Kossakowski, H.~Nastoll, A.~Bussiere, J.P.
  Guillaud, Phys. Lett. B \textbf{349}, 427 (1995)

\bibitem{Abele:1997}
H.~Abele, S.~Baeßler, D.~Dubbers, J.~Last, U.~Mayerhofer, C.~Metz, T.M.
  Müller, V.~Nesvizhevsky, C.~Raven, O.~Schärpf et~al., Phys. Lett. B
  \textbf{407}, 212 (1997)

\bibitem{Yerozolimsky:1997}
B.~Yerozolimsky, I.~Kuznetsov, Y.~Mostovoy, I.~Stepanenko, Phys. Lett. B
  \textbf{412}, 240 (1997)

\bibitem{Serebrov:1998}
A.P. Serebrov, I.A. Kuznetsov, I.V. Stepanenko, A.V. Aldushchenkov, M.S.
  Lasakov, Y.A. Mostovoi, B.G. Erozolimskii, M.S. Dewey, F.E. Wietfeldt,
  O.~Zimmer et~al., JETP \textbf{86}, 1074 (1998)

\bibitem{Abele:2002}
H.~Abele, M.~Astruc~Hoffmann, S.~Baeßler, D.~Dubbers, F.~Glück, U.~Müller,
  V.~Nesvizhevsky, J.~Reich, O.~Zimmer, Phys. Rev. Lett \textbf{88}, 211801
  (2002)

\bibitem{Soldner:2004}
T.~Soldner, L.~Beck, C.~Plonka, K.~Schreckenbach, O.~Zimmer, Phys. Lett. B
  \textbf{581}, 49 (2004)

\bibitem{Kreuz:2005}
M.~Kreuz, T.~Soldner, S.~Baeßler, B.~Brand, F.~Glück, U.~Mayer, D.~Mund,
  V.~Nesvizhevsky, A.~Petoukhov, C.~Plonka et~al., Phys. Lett. B \textbf{619},
  263 (2005)

\bibitem{Schumann:2007}
M.~Schumann, T.~Soldner, M.~Deissenroth, F.~Glück, J.~Krempel, M.~Kreuz,
  B.~Märkisch, D.~Mund, A.~Petoukhov, H.~Abele, Phys. Rev. Lett. \textbf{99},
  191803 (2007)

\bibitem{Schumann:2008}
M.~Schumann, M.~Kreuz, M.~Deissenroth, F.~Glück, J.~Krempel, B.~Märkisch,
  D.~Mund, A.~Petoukhov, T.~Soldner, H.~Abele, Phys. Rev. Lett. \textbf{100},
  151801 (2008)

\bibitem{Chupp:2012}
T.E. Chupp, R.L. Cooper, K.P. Coulter, S.J. Freedman, B.K. Fujikawa,
  A.~García, G.L. Jones, H.P. Mumm, J.S. Nico, A.K. Thompson et~al., Phys.
  Rev. C \textbf{86}, 035505 (2012)

\bibitem{Kozela:2012}
A.~Kozela, G.~Ban, A.~Białek, K.~Bodek, P.~Gorel, K.~Kirch, S.~Kistryn,
  O.~{Naviliat-Cuncic}, N.~Severijns, E.~Stephan et~al., Phys. Rev. C
  \textbf{85}, 045501 (2012)

\bibitem{Mund:2013}
D.~Mund, B.~Märkisch, M.~Deissenroth, J.~Krempel, M.~Schumann, H.~Abele,
  A.~Petoukhov, T.~Soldner, Phys. Rev. Lett. \textbf{110}, 172502 (2013)

\bibitem{Darius:2017}
G.~Darius, W.A. Byron, C.R. DeAngelis, M.T. Hassan, F.E. Wietfeldt, B.~Collett,
  G.L. Jones, M.S. Dewey, M.P. Mendenhall, J.S. Nico et~al., Phys. Rev. Lett.
  \textbf{119}, 042502 (2017)

\bibitem{Markisch:2018}
B.~M\"arkisch, H.~Mest, H.~Saul, X.~Wang, H.~Abele, D.~Dubbers, M.~Klopf,
  A.~Petoukhov, C.~Roick, T.~Soldner et~al., Phys. Rev. Lett. \textbf{122},
  242501 (2019)

\bibitem{baldoceolin:1994}
M.~Baldo-Ceolin, P.~Benetti, T.~Bitter, F.~Bobisut, E.~Calligarich, R.~Dolfini,
  D.~Dubbers, P.~El-Muzeini, M.~Genoni, D.~Gibin et~al., Zeitschrift f{\"u}r
  Physik C Particles and Fields \textbf{63}, 409 (1994)

\bibitem{Serebrov:2008a}
A.P. Serebrov, E.B. Aleksandrov, N.A. Dovator, S.P. Dmitriev, A.K. Fomin,
  P.~Geltenbort, A.G. Kharitonov, I.A. Krasnoschekova, M.S. Lasakov, A.N.
  Murashkin et~al., Phys. Lett. B \textbf{663}, 181 (2008)

\bibitem{Altarev:2009}
I.~Altarev, C.A. Baker, G.~Ban, K.~Bodek, M.~Daum, P.~Fierlinger,
  P.~Geltenbort, K.~Green, M.G.D. van~der Grinten, E.~Gutsmiedl et~al., Phys.
  Rev. D \textbf{80}, 032003 (2009)

\bibitem{Sun:2018}
X.~Sun, E.~Adamek, B.~Allgeier, M.~Blatnik, T.J. Bowles, L.J. Broussard, M.A.P.
  Brown, R.~Carr, S.~Clayton, C.~{Cude-Woods} et~al., Phys. Rev. C \textbf{97},
  052501 (2018)

\bibitem{Steyerl:1986}
A.~Steyerl, H.~Nagel, F.X. Schreiber, K.A. Steinhauser, R.~Gähler, W.~Gläser,
  P.~Ageron, J.~Astruc, W.~Drexel, G.~Gervais et~al., Phys. Lett. A
  \textbf{116}, 347  (1986)

\bibitem{Golub:1977}
R.~Golub, J.M. Pendlebury, Phys. Lett. A \textbf{62}, 337 (1977)

\bibitem{Zimmer:2007}
O.~Zimmer, K.~Baumann, M.~Fertl, B.~Franke, S.~Mironov, C.~Plonka, D.~Rich,
  P.~{Schmidt-Wellenburg}, H.F. Wirth, B.~{van den Brandt}, Phys. Rev. Lett.
  \textbf{99}, 104801 (2007)

\bibitem{Piegsa:2014}
F.M. Piegsa, M.~Fertl, S.N. Ivanov, M.~Kreuz, K.K.H. Leung,
  P.~{Schmidt-Wellenburg}, T.~Soldner, O.~Zimmer, Phys. Rev. C \textbf{90},
  015501 (2014)

\bibitem{Serebrov:1995}
A.P. Serebrov, {et al.}, Pis'ma Zh. Eksp. Teor. Fiz. \textbf{10}, 764 (1995)

\bibitem{Pattie:2016}
R.~Pattie, {LANL-UCN Team Team}, \emph{Commissioning of the Upgraded Ultracold
  Neutron Source at {{Los Alamos Neutron Science Center}}}, in \emph{{{APS
  Division}} of {{Nuclear Physics Meeting}}} (2016), p. DJ.005

\bibitem{Anghel:2009}
A.~Anghel, F.~Atchison, B.~Blau, B.~{van den Brandt}, M.~Daum, R.~Doelling,
  M.~Dubs, P.A. Duperrex, A.~Fuchs, D.~George et~al., Nucl. Instrum. Meth.
  Phys. Res. A \textbf{611}, 272 (2009)

\bibitem{Bison:2017}
G.~Bison, M.~Daum, K.~Kirch, B.~Lauss, D.~Ries, P.~{Schmidt-Wellenburg},
  G.~Zsigmond, T.~Brenner, P.~Geltenbort, T.~Jenke et~al., Phys. Rev. C
  \textbf{95}, 045503 (2017)

\bibitem{Abele_2012}
H.~Abele, H.~Leeb, New J. Phys. \textbf{14}, 055010 (2012)

\bibitem{Nesvizhevsky:2010}
V.V. Nesvizhevsky, A.Y. Voronin, R.~Cubitt, K.V. Protasov, Nature Physics
  \textbf{6}, 114 (2010)

\bibitem{Abele:2010}
H.~Abele, T.~Jenke, H.~Leeb, J.~Schmiedmayer, Phys. Rev. D \textbf{81}, 065019
  (2010)

\bibitem{Jenke:2011}
T.~Jenke, P.~Geltenbort, H.~Lemmel, H.~Abele, Nature Physics \textbf{7}, 468
  (2011)

\bibitem{Jenke:2014}
T.~Jenke, G.~Cronenberg, J.~Burgdörfer, L.A. Chizhova, P.~Geltenbort, A.N.
  Ivanov, T.~Lauer, T.~Lins, S.~Rotter, H.~Saul et~al., Phys. Rev. Lett.
  \textbf{112}, 151105 (2014)

\bibitem{Jenke:2018}
T.~Jenke, {et al.}, \emph{Testing {{Gravity}} at {{Short Distances}}: {{Gravity
  Resonance Spectroscopy}} with {{qBounce}}}, in \emph{Proceedings of
  {{PPNS}}-2018} (ILL Grenoble, France, 2018)
\end{thebibliography}
\end{document}